\documentclass[prd,aps,showpacs,nofootinbib,preprint,tightenlines,superscriptaddress]{revtex4}

\newcommand{\comment}[1]{}

\newcommand{\beq}{\begin{equation}}
\newcommand{\eeq}{\end{equation}}
\newcommand{\bqa}{\begin{eqnarray}}
\newcommand{\eqa}{\end{eqnarray}}

\begin{document}

\title{Hard-Loop Effective Action for Anisotropic Plasmas}

\preprint{ TUW-04-08 }

\author{Stanis\l aw Mr\'owczy\'nski}
\affiliation{So\l tan Institute for Nuclear Studies \\
ul. Ho\.za 69, PL - 00-681 Warsaw, Poland \\
and Institute of Physics, \'Swi\c etokrzyska Academy \\
ul. \'Swi\c etokrzyska 15, PL - 25-406 Kielce, Poland \\[5pt]
}
\author{Anton Rebhan}
%\author{Paul Romatschke}
\author{Michael Strickland}
\affiliation{Institut f\"ur Theoretische Physik, Technische Universit\"at Wien,
        Wiedner Hauptstrasse 8-10, A-1040 Vienna, Austria \vspace{0.5cm} }

%\date{March 23, 2004}

\vspace{0.3cm}

\begin{abstract}

We generalize the hard-thermal-loop effective action of the equilibrium
quark-gluon plasma to a non-equilibrium system which is space-time 
homogeneous but for which the parton momentum distribution is anisotropic. 
We show that the manifestly gauge-invariant Braaten-Pisarski form
of the effective action can be straightforwardly generalized and we
verify that it then generates
all $n$-point functions following from collisionless
gauge-covariant transport theory for a homogeneous anisotropic plasma.
On the other
hand, the Taylor-Wong form of the hard-thermal-loop effective action
has a more complicated generalization to the anisotropic case. Already
in the simplest case of anisotropic distribution functions, it
involves an additional term that is gauge invariant by itself, but
nontrivial also in the static limit.
\end{abstract}
\maketitle
\newpage

\small

%%%%%%%%%%%%%%%%%%%%%%%%%%%%%%%%%%%%%%%%%%%%%%%%%%%%%%%%%%%%%%%%%%%%%%%%%

\section{Introduction}

%%%%%%%%%%%%%%%%%%%%%%%%%%%%%%%%%%%%%%%%%%%%%%%%%%%%%%%%%%%%%%%%%%%%%%%%%

The hard thermal loop (HTL) approach \cite{Frenkel:br,HTLreviews}
has proved to be a crucial tool in describing 
the equilibrium quark-gluon plasma.  In particular it is absolutely
necessary for computing equilibrium and near-equilibrium quantities in a 
manner which is systematic and gauge independent.
However, we are often interested in non-equilibrium plasmas as in the 
case of relativistic heavy-ion collisions where a non-equilibrium parton 
system is expected to emerge during the early stages of the collision. To understand how 
the plasma evolves and thermalizes one has to go beyond the equilibrium 
description.  In this paper we focus on a specific non-equilibrium 
configuration which is (at least approximately) homogeneous and stationary 
but anisotropic in momentum space. Such an anisotropic quark-gluon plasma 
appears to be qualitatively different from the isotropic one as the 
quasi-particle collective modes can then be unstable 
\cite{Mrowczynski:xv,Mrowczynski:1996vh,Randrup:2003cw,Romatschke:2003ms,Arnold:2003rq}. 
And the presence of these instabilities can dramatically influence the 
system's evolution leading, in particular, to its faster equilibration. 

The gluon polarization tensor of a homogeneous and stationary but anisotropic 
plasma has been derived within semiclassical transport theory 
\cite{Mrowczynski:2000ed,Romatschke:2003ms} and diagrammatically 
\cite{Mrowczynski:2000ed}, following the formal rules of the HTL approach,
and the two approaches have been found to agree.
The anisotropic quark self-energy has been derived so far only 
diagrammatically \cite{Mrowczynski:2000ed,Arnold:2002zm}. 
However, the derivation is also possible within transport theory as it 
has been done in \cite{Blaizot:1993be} 
for the equilibrium plasma. 
The two-point 
functions - the gluon polarization and quark self energy - are sufficient 
to obtain, in particular, the spectrum of quasi-particles
and of unstable modes in the linear regime. However, one often 
needs the $n-$point functions to, for example, go beyond the lowest 
order of perturbative expansion. In the presence of instabilities,
soft $n-$point functions will be of importance to
the nonlinear phenomenon of saturation of instabilities, if the latter
is predominantly through interactions among the soft modes.

For the equilibrium plasma, the effective action, which 
summarizes the infinite set of hard thermal loop $n-$point functions, 
was first derived by Taylor and Wong \cite{Taylor:ia}, see
also \cite{Frenkel:ts,Efraty:1992gk}, 
and then a very elegant form was found by Braaten 
and Pisarski \cite{Braaten:1991gm}. The HTL effective action was also 
rederived within semiclassical transport theory 
\cite{Blaizot:1993be,Kelly:1994dh},
see also \cite{Manuel:2002pb}. 
The aim of this paper is to generalize the result 
to a non-equilibrium system which is space-time homogeneous but anisotropic 
in momentum space. (We call it the `hard loop action'; the word `thermal'
is dropped as it refers to equilibrium.) 

We show that the HTL effective action as written down by
Braaten and Pisarski \cite{Braaten:1991gm} generalizes naturally
to the anisotropic case. 
We verify that this more general hard-loop effective action
is still equivalent to gauge-covariant semiclassical transport theory
\cite{Blaizot:1993be}. On the other hand, the HTL effective action
in the form of Taylor and Wong \cite{Taylor:ia} has a more
complicated generalization for anisotropic plasmas. In addition
to the structure which is present in the equilibrium case and
which has a ``secret'' Chern-Simons nature \cite{Efraty:1992gk},
there are additional manifestly gauge-invariant
contributions which have a nontrivial static limit.
Finally, we derive explicit expressions for the quark-gluon, triple-gluon, and
four-gluon vertices for an anisotropic system, verify that
they satisfy the appropriate Ward-Takahashi identities, and compare their integral
representations with those of the isotropic case.

%%%%%%%%%%%%%%%%%%%%%%%%%%%%%%%%%%%%%%%%%%%%%%%%%%%%%%%%%%%%%%%%%%%%%%%

\section{Effective action}

%%%%%%%%%%%%%%%%%%%%%%%%%%%%%%%%%%%%%%%%%%%%%%%%%%%%%%%%%%%%%%%%%%%%%%%

To construct the effective action we will first find a form 
which can generate the anisotropic gluon 
polarization tensor and quark self-energy which have been obtained
in previous works \cite{Mrowczynski:2000ed,Romatschke:2003ms,Arnold:2003rq}.
We will then use the requirement of gauge invariance to extend
the result to the full effective action for quarks and gluons.

%\subsection{Gluon polarization tensor}

The anisotropic gluon polarization tensor derived in 
\cite{Mrowczynski:2000ed,Romatschke:2003ms} can be written 
in momentum space as
\bqa \label{g-self}
\Pi^{\mu \nu}_{ab}(k) = \delta_{ab} { g^2 \over 2 } \int_{\bf p} 
{ f({\bf p}) \over |{\bf p}| } 
{ (p\cdot k)(k^\mu p^\nu + p^\mu k^\nu) - k^2 p^{\mu} p^{\nu}
- (p\cdot k)^2 g^{\mu\nu} \over(p\cdot k)^2} \;,
\eqa
\comment{m}
where $\mu, \nu$ denote Lorentz indices and $a, b$ color indices 
in adjoint representation; $g$ is the coupling constant and 
$$
 \int_{\bf p} \cdots \equiv \int {d^3 p \over (2\pi)^3} \cdots\Big|_{p_0=|\mathbf p|} \;.
$$
\comment{m}
The distribution function $f({\bf p})$ in Eq.~(\ref{g-self}) is the effective parton momentum distribution which describes 
partons (quarks and gluons) which are on mass-shell.  We assume that it
only depends on three-momentum and is independent of the spatial coordinates
(homogeneous) and therefore has the form
\bqa
f({\bf p}) \equiv 2 N_f \left(n({\bf p}) + \bar n ({\bf p})\right) + 4 N_c n_g({\bf p}) \; ,
\eqa
\comment{m}
where $n$, $\bar n$, and $n_g$ are the distribution functions of quarks,
antiquarks and gluons. In equilibrium these distribution functions
reduce to the standard Fermi-Dirac and Bose-Einstein distributions
\bqa \label{equilibrium}
n^{\rm eq}({\bf p}) &=& {1 \over \exp(|{\bf p}|  - \mu)/T+1} \;,
\nonumber \\
\bar n^{\rm eq}({\bf p}) &=& {1 \over \exp(|{\bf p}|  + \mu)/T+1} \;,
\nonumber \\
n_g^{\rm eq}({\bf p}) &=& {1 \over \exp (|{\bf p}|/T) - 1} \;,
\eqa
\comment{m}
with $T$ and $\mu$ denoting the temperature and chemical potential
and both quarks and gluons are assumed to be massless. We note the gluon 
self energy in the form (\ref{g-self}) is explicitly Lorentz covariant, 
symmetric with respect to the Lorentz indices and transversal
($k_{\mu} \Pi^{\mu \nu}(k)=0$). 

%\subsection{Quark Self-energy}

The quark self energy for an anisotropic system has been
obtained previously \cite{Mrowczynski:2000ed} and is 
given by
\bqa \label{q-self}
\Sigma(k) = {C_F \over 4} g^2 
\int_{\bf p} { \tilde f ({\bf p}) \over |{\bf p}|} 
{p \cdot \gamma \over p\cdot k} \;, 
\eqa
\comment{m}
where $C_F \equiv (N_c^2 -1)/2N_c$ and  
$$
\tilde f ({\bf p}) \equiv 2 \left( n({\bf p}) + \bar n ({\bf p}) \right) + 4 n_g({\bf p}) \; .
$$
%

%\subsection{Effective action}

We now attempt to find an action which can generate the
anisotropic gluon polarization tensor (\ref{g-self}) and quark
self-energy (\ref{q-self}).
The corresponding terms in the action will have the form
\bqa \label{g-2-action1}
{\cal L}^{(A)}_2(x) &=&  
{1\over2} \int_y A^a_\mu(x) \Pi^{\mu \nu}_{ab}(x-y) A^b_\nu(y) \;, 
\\ [2mm]  \label{q-2-action1}
{\cal L}^{(\Psi)}_2(x) &=&  
\int_y \bar{\Psi}(x) \Sigma (x-y) \Psi (y) \; ,
\eqa
\comment{m}
where
$$
\int_y \cdots \equiv \int d^4y \cdots \;;
$$
and the subscript `2' indicates that the effective actions above only
generate two-point functions.  These actions will then be extended
to generate all $n$-point functions by writing them in a gauge 
invariant form.

%\subsection{Quark effective action}
 
Using the explicit form of the quark self energy (\ref{q-self}),
one immediately rewrites the action (\ref{q-2-action1}) as
\bqa \label{q-2-action2}
{\cal L}^{(\Psi)}_2(x) =  
- i {C_F \over 4}  g^2 \int_{\bf p} { \tilde f ({\bf p}) \over |{\bf p}|}
\; \bar{\Psi}(x) {p \cdot \gamma \over p\cdot \partial}
\Psi (x) \; ,
\eqa
where
$$
{1 \over p\cdot \partial} \Psi (x) \equiv
i \int_k {e^{-ikx} \over p\cdot k} \: \Psi (k) \;.
$$ 
Following Braaten and Pisarski \cite{Braaten:1991gm}, we modify the 
action (\ref{q-2-action2}) to comply with the requirement of gauge invariance.
We simply replace the derivative $\partial^\mu$ by the covariant
derivative $D^\mu = \partial^\mu - ig A^\mu$ in the fundamental
representation. Thus, we obtain
\bqa \label{q-action}
{\cal L}^{(\Psi)}(x) = - i {C_F \over 4} g^2 
\int_{\bf p} { \tilde f ({\bf p}) \over |{\bf p}|}
\; \bar{\Psi}(x) {p \cdot \gamma \over p\cdot D}
\Psi (x) \;.
\eqa
Note that when expanding the covariant derivative in the denominator above
one needs to take care about the ordering of the fields and operators.
\beq 
{1 \over p\cdot D}\: \Psi (x)  \buildrel \rm def \over =
{1\over p\cdot \partial}
\sum_{n=0}^{\infty}\bigg(ig \:  p\cdot A(x) {1 \over p \cdot \partial} \bigg)^n
\: \Psi (x) \;, 
\eeq
so that, for example, the first and second order expansions are
\beq
{ig \: p \cdot A(x) \over p \cdot \partial} \: \Psi (x)
\equiv 
- g \: p \cdot A(x) \int_k {e^{-ikx} \over p\cdot k} \: \Psi (k) \;,
\eeq
and
\beq
\bigg({ig \: p \cdot A(x) \over p \cdot \partial} \bigg)^2 \Psi (x)
\equiv
 g^2 p \cdot A(x) \int_q  {e^{-iqx} \over p\cdot q} 
 \int_{x'}  e^{iqx'} \:
 p \cdot A(x') \int_k {e^{-ikx'} \over p\cdot k} \: \Psi (k) \;.
 \eeq

In equilibrium, where the quark and gluon distribution functions are 
given by Eqs.~(\ref{equilibrium}), the integrals over the momentum length
and over the angle factorize, and the action (\ref{q-action}) reduces
to the Braaten-Pisarski result
$$
{\cal L}^{(\Psi)}_{\rm HTL}(x) =  
- i m_q^2  \, \Big\langle \bar{\Psi}(x) {\hat p \cdot \gamma \over \hat p\cdot D}
\Psi (x) \Big\rangle_{\hat{\bf p}} \;,
$$ 
\comment{m}
where
$$
m_q^2 = {C_F \over 4} g^2 
\int_{\bf p} { \tilde f ^{\rm eq}({\bf p}) \over |{\bf p}|}
= {C_F \over 8} g^2 \bigg(T^2 + {\mu^2 \over \pi^2} \bigg) \;,
$$
\comment{m}
and $\langle \cdots \rangle_{\hat{\bf p}} \equiv \int {d^2\Omega \over 4\pi} \cdots\;\;$ 
denotes an average over the orientation of the unit vector 
$\hat{\bf p} = {\bf p}/ |{\bf p}|$ which defines the four-vector 
$\hat p \equiv (1, \hat{\bf p})$.

%\subsection{Gluonic effective action}

Let us now discuss the gluon effective action. At first, we look 
for an operator ${\cal M}^{\mu \nu}(x)_{ab}$ that satisfies
the equation
$$
{1 \over 2} \int_y A^a_\mu(x) \Pi^{\mu \nu}_{ab}(x-y) A^b_\nu(y) 
= {1\over 4} (\partial_\mu A_\nu^a (x) - \partial_\nu A_\mu^a (x))
{\cal M}^{\nu \rho}_{ab}(x) 
(\partial_\rho A^{b \,\mu} (x) - \partial^\mu A_\rho^b (x)) \;,
$$
giving
\bqa \label{Pi-M}
\Pi^{\mu \nu}_{ab}(k) = - 2 k^2 {\cal M}^{\sigma \rho}_{ab}(k)
\; P_{\rho \sigma}^{\;\;\;\;\mu \nu}(k) \;,
\eqa
where
$$
P^{\rho \sigma \mu \nu}(k) =
{1 \over k^2} \Big[ k^2 g^{\rho \nu} g^{\sigma \mu} 
+ k^\rho k^\sigma g^{\mu \nu} 
- k^\rho k^\nu g^{\sigma \mu}
- k^\sigma k^\mu  g^{\rho \nu} \Big] \;.
$$ 
Since $P$ is the projection operator
($P^{\rho \sigma \mu \nu}(k) P_{\nu \mu}^{\;\;\;\; \delta \lambda}(k) 
= - P^{\rho \sigma  \delta \lambda}(k)$), $P^{-1}$ does not exist.
Therefore, there is no unique solution of Eq.~(\ref{Pi-M}); various 
solutions differ from each other by the components parallel to $k$. 
Because 
$k_\mu P^{\rho \sigma \mu \nu}(k) = k_\nu P^{\rho \sigma \mu \nu}(k) = 0$,
Eq.~(\ref{Pi-M}) complies with the transversality of $\Pi^{\mu \nu}(k)$.

Substituting the explicit form of the gluon self energy (\ref{g-self})
in Eq.~(\ref{Pi-M}), one finds that the equation is satisfied by
$$
{\cal M}^{\mu \nu}_{ab}(k) = - \delta_{ab} \: {g^2 \over 2} \int_{\bf p}
{ f({\bf p}) \over |{\bf p}| } \; 
{p^\mu p^\nu \over (p \cdot k)^2} \;,
$$
which gives
\bqa \label{g-2-action2}
{\cal L}^{(A)}_2(x) =
- {g^2\over2} \int_{\bf p}
{ f({\bf p}) \over |{\bf p}| } \;  
(\partial_\mu A_\nu^a (x) - \partial_\nu A_\mu^a (x))
{p^\nu p^\rho \over (p \cdot \partial)^2} \;
(\partial_\rho A^{a\,\mu} (x) - \partial^\mu A_\rho^a (x))  \;.
\eqa
In order to generate the higher-order vertices we invoke the
requirement of gauge invariance, replacing
$\partial^\mu A^\nu_a - \partial^\nu A^\mu_a $ by the field strength
tensor
$F^{\mu \nu}_a \equiv 
\partial^\mu A^\nu_a - \partial^\nu A^\mu_a + g f_{abc} A^\mu_b A^\nu_c$,
and $\partial^\mu$ by the covariant derivative in the adjoint representation 
$D^\mu_{ab} \equiv \partial^\mu \delta_{ab} + g f_{acb} A^\mu_c$. Thus, 
we obtain the effective action
\bqa \label{g-action}
{\cal L}^{(A)}(x) =
- {g^2\over2} \int_{\bf p}
{ f({\bf p}) \over |{\bf p}| } \;  
F_{\mu \nu}^a (x)
\bigg({p^\nu p^\rho \over (p \cdot D)^2} \bigg)_{ab} \;
F_\rho^{\;\;b \,\mu} (x) \;.
\eqa
\comment{m}

In equilibrium, the gluon action (\ref{g-action}) reduces, as the 
quark action, to the respective Braaten-Pisarski result
$$
{\cal L}^{(A)}_{\rm HTL}(x) = 
- m_\infty^2
\Big\langle 
F_{\mu \nu}^a (x)
\bigg({\hat p^\nu \hat p^\rho \over (\hat p \cdot D)^2} \bigg)_{ab} \;
F_\rho^{\;\;b \,\mu} (x) 
\Big\rangle_{\hat{\bf p}} \;,
$$
\comment{m}
where
$$
m_\infty^2 = {g^2\over2} \int_{\bf p} 
{ f^{\rm eq}({\bf p}) \over |{\bf p}| } 
= {N_c \over 6} g^2 T^2 
+ {N_f \over 12} g^2 \bigg(T^2 + {3 \over \pi^2} \mu^2 \bigg) \;.
$$
\comment{m}
%

%\subsection{Total Effective Action}

To summarize, the generalization of the HTL effective action of
Braaten and Pisarski to the anisotropic case is simply given by
\begin{equation}\label{Saniso}
  S_{\rm aniso}=
-{g^2\over2} \int_x \int_{\mathbf p} %{d^4 p\, \delta^{(+)}(p) \over (2\pi)^3}
\left\{ { f(\mathbf p) \over |{\bf p}| }\;
F_{\mu \nu}^a (x)
\bigg({p^\nu p^\rho \over (p \cdot D)^2} \bigg)_{ab} \;
F_\rho^{\;\;b \,\mu} (x)
+ i {C_F \over 2}
{ \tilde f ({\bf p}) \over |{\bf p}| }\;
 \bar{\Psi}(x) {p \cdot \gamma \over p\cdot D}
\Psi (x) \right\}.
\end{equation}
%where we have rewritten $\int_{\bf p}\cdots|_{p_0=|\mathbf p|}$
%with the help of 
%$\delta^{(+)}(p)=\theta(p_0)\delta(p^2)$
%as a four-dimensional momentum integral.

%%%%%%%%%%%%%%%%%%%%%%%%%%%%%%%%%%%%%%%%%%%%%%%%%%%%%%%%%%%%%%%%%%%%%%%

\section{Equivalence with gauge-covariant kinetic theory}

%%%%%%%%%%%%%%%%%%%%%%%%%%%%%%%%%%%%%%%%%%%%%%%%%%%%%%%%%%%%%%%%%%%%%%%

The hard loop effective action (\ref{Saniso}) is manifestly
gauge invariant and it contains the two-point functions obtained
previously from gauge-covariant transport equations \cite{Mrowczynski:2000ed}. 
Hence, it
is a good candidate for generating all of the hard-loop vertex functions
of a gauge-covariant kinetic theory. 
That this is indeed the case is not entirely obvious, at least
for the gauge-boson part of the effective action, since the latter
contains higher powers of inverse gauge-covariant line derivatives
than is suggested by the structure of the kinetic equations.
Fortunately, however, the proof of equivalence that has been
worked out in detail in Ref.~\cite{Blaizot:1993be}, can be shown
to carry over almost line by line as long as the distribution
functions $f$ and $\tilde f$ are $x$-independent.

Vertex functions containing external fermion lines are generated
by the fermionic current $\eta=\delta S/\delta \bar\Psi$
and this is indeed of the same form as the fermionic current
one can define in gauge-covariant kinetic theory \cite{Blaizot:1993be}.
%,
%provided that all fields are restricted such that
%they do not have components in the kernel of the
%covariant line derivative \cite{Blaizot:1993be}.
The generalization of this proof of equivalence
to anisotropic distributions functions $\tilde f$
in the fermionic effective
action (\ref{q-action}) is trivial since $\tilde f(\mathbf p)$ appears
in undifferentiated form in either formalism (see the
appendix of Ref.~\cite{Blaizot:1993be}).

Vertex functions containing only external gauge-boson fields
can be obtained by expanding the induced current
$j^\mu[A]$ in powers of the gauge field $A^\mu$.
Solving the gauge-covariant transport equations in the isotropic
\cite{Blaizot:1993be} as well as in the anisotropic case
\cite{Mrowczynski:2000ed} yields an induced current of the
form
\begin{equation}\label{Jind0}
j^\mu[A] = -g^2
\int {d^4p\over(2\pi)^3} 
\delta^{(+)}(p) \,p^\mu\, {\partial f(\mathbf p) \over \partial p_\beta}
[p\cdot D(A)]^{-1} F_{\beta\gamma}(A) p^\gamma \, ,
\end{equation}
where for emphasis we have written out $\int_{\mathbf p}$
as a four-dimensional momentum integral with 
$\delta^{(+)}(p)\equiv\theta(p_0)\delta(p^2)$.

The hard-loop effective action (\ref{Saniso}), on the other hand,
involves an undifferentiated distribution function $f(\mathbf p)$, 
so as a first
step we should partially integrate the derivative with respect to $p$.
This is in fact possible without picking up contributions from
the integration measure, because differentiating $\delta(p^2)$
would produce $p^\beta$, but $p^\beta p^\gamma F_{\beta\gamma}(A) \equiv 0$.
Also, differentiating $\theta(p_0)$ is harmless if 
$\lim_{\mathbf p\to0} \mathbf p^2 f(\mathbf p)=0$,
since it involves
$$\int d\Omega_{\hat \mathbf p} \int_0^\infty d|\mathbf p| 
\delta(|\mathbf p|)|\mathbf p|^2f(\mathbf p)\{\hat \mathbf p^i
[\hat \mathbf p\cdot \mathbf D(A)]^{-1} F_{0j}(A) \hat \mathbf p^j\} \, , $$
with $\hat \mathbf p^i=\mathbf p^i/|\mathbf p|$.
We can therefore write
\begin{equation}\label{Jind}
j^\mu[A] = g^2
\int_{\mathbf p} % {d^4p \over (2\pi)^3} \delta^{(+)}(p) 
f(\mathbf p)
{\partial\over\partial p_\beta}
\{p^\mu [p\cdot D(A)]^{-1} F_{\beta\gamma}(A) p^\gamma\}.
\end{equation}
From this form one can immediately infer that
this induced current is covariantly
conserved, 
$$D[A]\cdot J[A]\propto
\int_{\mathbf p} % d^4p \delta^{(+)}(p) 
f(\mathbf p)
F_{\beta\gamma}(A) g^{\beta\gamma} \equiv 0.
$$
This implies that an effective action from which this induced
current can be derived according to $j^\mu=\delta S/\delta A_\mu$
must be gauge invariant, since gauge invariance is equivalent
to $D[A]\delta S/\delta A \equiv 0$ (which further differentiated
gives all the Ward identities).

In the form (\ref{Jind}), the induced current is indeed exactly
analogous to the HTL case for which Ref.~\cite{Blaizot:1993be}
has shown equivalence with the first functional derivative of
the Braaten-Pisarski effective action. The corresponding proof
is somewhat lengthy (see Eqs.~(C.15)--(C.27) of Ref.~\cite{Blaizot:1993be})
and we shall not repeat it here. 
It involves representing formal relations like
$$ [(p\cdot D)^{-1}, D^\beta]_{ab}
=((p\cdot D)^{-1} [ D^\beta,p\cdot D ] (p\cdot D)^{-1})_{ab}
=(p\cdot D)^{-1}_{ac} g f_{ced} F_e^{\beta\gamma}p_\gamma
(p\cdot D)^{-1}_{db}
$$
in terms of gauge-covariant parallel transporters.
The essential point to notice
is that once $j^\mu[A]$ is expressed in terms of an undifferentiated
distribution function, the remaining steps are independent of
the form $f(\mathbf p)$ as long as it is homogeneous in $x$-space.

Another point that should be noted is that the equivalence
of the effective action with the kinetic equations strictly speaking
holds true only on the space of fields $\mathcal R$
where all gauge-covariant
line derivatives $p\cdot D(A)$ have vanishing kernel and can be
inverted without regard of boundary conditions \cite{Blaizot:1993be}. 
%Physically this
%corresponds to restricting oneself to field components which do not
%experience Landau damping. 
For unrestricted fields it is only
at the level of vertex functions or kinetic equations that the
formal expressions become well-defined, because only then one can
impose specific boundary conditions.

% (on the space of fields
%with vanishing kernel of gauge-covariant line derivatives),
%see Eqs.~(C.15)--(C.27) of Ref.~\cite{Blaizot:1993be}.

%%%%%%%%%%%%%%%%%%%%%%%%%%%%%%%%%%%%%%%%%%%%%%%%%%%%%%%%%%%%%%%%%%%%%%%

\section{Taylor-Wong form}

%%%%%%%%%%%%%%%%%%%%%%%%%%%%%%%%%%%%%%%%%%%%%%%%%%%%%%%%%%%%%%%%%%%%%%%

Originally, the HTL effective action was obtained by Taylor and
Wong \cite{Taylor:ia} in a form which is not manifestly gauge invariant, but
involves only a single power of inverse gauge-covariant line derivatives.
The Taylor-Wong form has also the advantage of making it evident
that all higher-point HTL vertex functions vanish in the static
limit, and that the two-point functions then
reduce to a simple momentum-independent
mass term.

Explicit calculations of the two-point functions have shown that this
simplicity of the static limit does not carry over to the anisotropic
case \cite{Romatschke:2003ms,Birse:2003qp}. 
However, it is instructive to see explicitly where
anisotropic distributions functions spoil the equivalence of the
Braaten-Pisarski form (which does easily generalize to the anisotropic
case) with the Taylor-Wong form (which evidently does not).
To this end, we start by rewriting the induced current in the form
of Eq.~(\ref{Jind0}) as
\begin{equation}
  j^\mu[A]=-g^2\int_{\mathbf p} %{d^4p\over(2\pi)^3} \delta^{(+)}(p) 
p^\mu {\partial f(\mathbf p) \over \partial p_\beta}
{1\over p\cdot D}\left( F_{\beta0}\, p^0+F_{\beta i}\, p^i
\right).
\end{equation}

In the isotropic case one has $\partial f(\mathbf p) / \partial p_\beta
\propto \delta^\beta_j p^j$, so the second term in the parenthesis vanishes
because $F_{ij}$ is antisymmetric,
whereas in the first one can use that
$F_{\beta0}=D_\beta A_0-\partial_0 A_\beta$ and $F_{00}\equiv 0$
so that
\begin{equation}
  j^\mu_{\rm iso}[A]=
-g^2\int_{\mathbf p} % {d^4p\over(2\pi)^3} \delta^{(+)}(p) 
{p^\mu\over|\mathbf p|}  f'(|\mathbf p|) 
\left(A_0-{1\over p\cdot D}\partial_0 (p\cdot A) \right)
,
\end{equation}
which is exactly the first functional derivative of the Taylor-Wong
effective action.

In the anisotropic case, these manipulations are clearly no longer
possible. Specialising to the case where $f$ depends on just
the energy $p_0=|\mathbf p|$ and a projection of $\mathbf p$ on
a fixed spatial direction $\mathbf n$, one can write
$${\partial f(\mathbf p) \over \partial p_\beta}
=\delta^\beta_j \left( {p^j\over p_0^2}f_1 + {n^j\over p_0}f_2 \right).$$
The induced current for the anisotropic case can then be decomposed
according to
\begin{equation}
j^\mu_{\rm aniso}[A]=-g^2\int_{\mathbf p} %{d^4p\over(2\pi)^3} \delta^{(+)}(p) 
{p^\mu\over|\mathbf p|}  \left\{
f_1\left(A_0-{1\over p\cdot D}\partial_0 (p\cdot A) \right)
+f_2 {1\over p\cdot D} n^j F_{j\nu}p^\nu \right\}
.
\end{equation}
In this form one has one contribution $\propto f_1$ which is exactly analogous
to the Taylor-Wong effective action. This part is gauge invariant
by itself, although its gauge invariance is not manifest, and it
reduces to a simple (constant) mass term for $A_0$ in the static limit.
On the other hand, the second part, which is specific to the 
anisotropic case ($f_2\not\equiv0$) is manifestly gauge invariant,
but it has nontrivial momentum-dependence even in the static limit,
and correspondingly generates nontrivial higher-point functions
also in the static limit.

%%%%%%%%%%%%%%%%%%%%%%%%%%%%%%%%%%%%%%%%%%%%%%%%%%%%%%%%%%%%%%%%%%%%%%%

\section{Vertex functions}

%%%%%%%%%%%%%%%%%%%%%%%%%%%%%%%%%%%%%%%%%%%%%%%%%%%%%%%%%%%%%%%%%%%%%%%

In this section we collect expressions for the quark-gluon, triple-gluon,
and four-gluon vertex functions for an anisotropic system.  We also show 
explicitly that these vertex functions satisfy the appropriate 
Ward-Takahashi identities.  As we have discussed previously the effective 
action (\ref{Saniso}) is gauge invariant by construction so that these
identities are guaranteed to be satisfied; however, due to the complexity
of the resulting vertex functions the explicit checks provide confidence
that the vertex functions derived are correct.

\subsection{Quark-Gluon Vertex function}

When the effective action (\ref{Saniso}) is expanded in powers of the
quark and gluon fields there appears a term of the form
$$
\int_y \int_z \bar{\Psi}(x) \: \Lambda^\mu (x,y,z) \: \Psi (y) \: A_\mu (z) \;,
$$
where $\Lambda^\mu (x,y,z)$ is quark-gluon the vertex function. 
To obtain this term we need only expand the action (\ref{q-action}) to
leading order in the gluon field strength 

\bqa
{\cal L}^{(\Psi)}(x) &=& - {i C_F \over 4} g^2 
\int_{\bf p} { \tilde f ({\bf p}) \over |{\bf p}|}
\; \bar{\Psi}(x) {p \cdot \gamma \over p\cdot D}
\Psi (x)  \nonumber \\
&=& - {i C_F \over 4} g^2 \int_{\bf p} { \tilde f ({\bf p}) \over |{\bf p}|}
\; \bar{\Psi}(x) \: {p \cdot \gamma  \over p\cdot \partial}
\sum_{n=0}^{\infty}\bigg( {i \,g \, p\cdot A(x) \over p \cdot \partial} \bigg)^n
\Psi (x)  
%\nonumber \\
%&=& - {i C_F \over 2} g^2 \int_{\bf p} { \tilde f ({\bf p}) \over |{\bf p}|}
%\; \bar{\Psi}(x) \: {p \cdot \gamma  \over p\cdot \partial}
%\left( 1 + { i \, g \, p \cdot A(x) \over p \cdot \partial} + {\cal O}(g^2) \right)
%\Psi (x) \; .
\eqa
After Fourier transformation the ${\cal O}(g^3)$ contribution 
above gives
\bqa 
\label{vertexdef}
\Lambda^\mu_a (q_1,q_2,k) = i g t^a \: (2\pi)^4 \delta^{(4)}(q_1 + q_2 + k) \;
\Lambda^\mu(q_1,q_2,k) \eqa
with
\bqa 
\label{vertex}
\Lambda^\mu(q_1,q_2,k) = {C_F \over 4} g^2
\int_{\bf p} { \tilde f ({\bf p}) \over |{\bf p}|} 
{\hat{p}\!\cdot\!\gamma \over \hat{p}\!\cdot\!q_1\;\hat{p}\!\cdot\!q_2} \: \hat{p}^\mu\;.
\eqa
where $q_1$ and $q_2$ are outgoing quark momentum and $k$ is the outgoing gluon
momentum.  
The matrix $t^a$ is in the fundamental representation
of the $SU(N_c)$ algebra with the standard normalization
${\rm tr}(t^a t^b) = {1 \over 2} \delta^{ab}$.
To verify that this vertex function (\ref{vertex}) obeys 
the Ward-Takahashi identity we contract it with the external gluon momentum
to obtain
\bqa 
\label{qgward}
k_\mu\Lambda^\mu (q_1,q_2,k) =  \Sigma (q_1) + \Sigma (q_2) \;,
\eqa
which is just the Ward-Takahashi identity.

%%%%%%%%%%%%%%%%%%%%%%%%%%%%%%%%%%%%%%%%%%%%%%%%%%%%%%%%%%%%%%%%%%%%%%%%5

\subsection{Triple-Gluon Vertex}

%%%%%%%%%%%%%%%%%%%%%%%%%%%%%%%%%%%%%%%%%%%%%%%%%%%%%%%%%%%%%%%%%%%%%%%%5

In order to obtain the triple-gluon coupling or gluon three-point vertex we have to expand the action (\ref{g-action}) to order $A^3$ to obtain all terms of the form
$$
\Gamma^{\mu\nu\lambda}(x,y,z) \: A_\mu(x) \: A_\nu(y) \: A_\lambda (z) \;,
$$
where $\Gamma^{\mu\nu\lambda}(x,y,z)$ is the triple-gluon vertex function. 

At this order there are two types of contributions.  One comes from terms which are of the
form $(\partial A) A A$ coming from the leading-order expansion of the kernel contracted
with the non-abelian part of the field strength tensor and the others are of the form
$(\partial A)^2 A$ coming from the next-to-leading order expansion of the covariant
derivative in the kernel contracted against the abelian part of the field strength
tensor.  The first type are given by
\bqa 
  {\cal L}_1 \sim 2 (\partial_\mu A^c_\alpha - \partial_\alpha A_\mu^c)
        {\cal T}^{\alpha \beta}(\partial) A^\mu_a A_\beta^b f^{abc} \; , 
        \eqa
and the second type are given by
\bqa
  {\cal L}_2 \sim 2 (\partial_\mu A^a_\alpha - \partial_\alpha A_\mu^a) 
        {\cal T}^{\alpha \beta}(\partial) A_\gamma^b {\cal T}^{\gamma}(\partial)
         (\partial^\mu A^c_\beta - \partial_\beta A^\mu_c) f^{abc} \; .
\eqa
where $f^{abc}$ are the $SU(N_c)$ structure
constants and we have introduced the $n$-tensor
\bqa
{\cal T}^{\mu_1 \mu_2 \cdots \mu_n}(\partial) = (p\cdot\partial)^{-n} \sum_{i=1}^{n} p^{\mu_i} \; ,
\eqa
which in momentum-space is defined by
\bqa
{\cal T}^{\mu_1 \mu_2 \cdots \mu_n}(k) = (p\cdot k)^{-n} \sum_{i=1}^{n} p^{\mu_i} \; .
\eqa
Note that these tensors are totally symmetric in all Lorentz indices and that
products of these tensors are also symmetric in the resulting indices, 
e.g. $ {\cal T}^\mu(k){\cal T}^\nu(q) = {\cal T}^\nu(k){\cal T}^\mu(q)$. 

We then Fourier transform the resulting expressions and relabel indices so that
all contributions are of the form of a three tensor contracted with
$A_\mu^a(k) A_\nu^b(q) A_\lambda^c(r) f^{abc}$ where $k, q, r$ are the incoming
gluon momentum which satisfy $k+q+r=0$.  This gives
$$
2 \left( (q\cdot r) {\cal T}^{\mu\nu}(r) {\cal T}^{\lambda}(q)
- {\cal T}^\mu(r) {\cal T}^\nu(q) q^\lambda \right) \, .
$$
From here we must sum over all
permutations of the sets $(k,\mu,a)$, $(q,\nu,b)$, and $(r,\lambda,c)$ taking
into account the minus signs coming from $f^{abc}$ whenever appropriate.
Defining 
\beq
\Gamma^{\mu\nu\lambda}_{abc}(k,q,r) = i g (2\pi)^4 \delta^{(4)}(k+q+r) f^{abc} \Gamma^{\mu\nu\lambda}(k,q,r) 
\eeq
we obtain
\bqa
\Gamma^{\mu\nu\lambda}(k,q,r) &=& 
{g^2 \over 2}
\int_{\bf p} { f({\bf p}) \over |{\bf p}| } \;  
\Biggl[
(k\cdot r) \left( 
{\cal T}^{\mu\nu}(k) {\cal T}^{\lambda}(r)
-{\cal T}^{\mu\nu}(r) {\cal T}^{\lambda}(k) \right)
\nonumber \\ &&
+(q\cdot k) \left( 
{\cal T}^{\mu\nu}(q) {\cal T}^{\lambda}(k)
-{\cal T}^{\mu\nu}(k) {\cal T}^{\lambda}(q) \right)
+(q\cdot r) \left( 
{\cal T}^{\mu\nu}(r) {\cal T}^{\lambda}(q)
-{\cal T}^{\mu\nu}(q) {\cal T}^{\lambda}(r) \right)
\nonumber \\ &&
- {\cal T}^\mu(r) {\cal T}^\nu(q) q^\lambda
+ {\cal T}^\mu(k) {\cal T}^\nu(r) k^\lambda
- {\cal T}^\mu(k) {\cal T}^\lambda(q) k^\nu
+ {\cal T}^\mu(q) {\cal T}^\lambda(r) r^\nu
\nonumber \\ &&
\hspace{4cm}
- {\cal T}^\nu(k) {\cal T}^\lambda(r) r^\mu
+ {\cal T}^\nu(q) {\cal T}^\lambda(k) q^\mu
\Biggr]\, .
\label{3gluevertex}
\eqa
Note that $\Gamma^{\mu\nu\lambda}(k,q,r)$ is totally symmetric in its three indices and traceless in any
pair of indices, e.g. $g_{\mu\nu}{\cal T}^{\mu\nu\lambda}=0$, and that
it is odd (even) under odd (even) permutations of the momenta $k$, $q$, and
$r$.
To verify that this vertex obeys the Ward-Takahashi identity we
contract with $k_\mu$ to obtain
\bqa
k_\mu \Gamma^{\mu\nu\lambda}(k,q,r) &=& 
{g^2 \over 2}
\int_{\bf p} { f({\bf p}) \over |{\bf p}| } \;  
\Biggl[ 
{\cal T}^{\lambda}(q) q^\nu + {\cal T}^{\nu}(q) q^\lambda - q^2 {\cal T}^{\nu\lambda}(q) - g^{\nu\lambda}
\nonumber \\
&&-{\cal T}^{\lambda}(r) r^\nu - {\cal T}^{\nu}(r) r^\lambda + r^2 {\cal T}^{\nu\lambda}(r) + g^{\nu\lambda}
\Biggr] \, ,
\eqa
When expressed in terms of the ${\cal T}$ tensors the gluon self-energy (\ref{g-self}) is
\bqa
\Pi^{\nu\lambda}(q) &=& 
{g^2 \over 2}
\int_{\bf p} { f({\bf p}) \over |{\bf p}| } \;  
\Biggl[
{\cal T}^{\lambda}(q) q^\nu + {\cal T}^{\nu}(q) q^\lambda - q^2 {\cal T}^{\nu\lambda}(q) - g^{\nu\lambda}
\Biggr] \, ,
\eqa
thus we can see that
\bqa
k_\mu \Gamma^{\mu\nu\lambda}(k,q,r) &=& 
\Pi^{\nu\lambda}(q) - \Pi^{\nu\lambda}(r) \, ,
\eqa
which is simply the Ward-Takahashi identity.

Note also that it is possible to simplify (\ref{3gluevertex}) by integrating by parts to obtain
\bqa
\Gamma^{\mu\nu\lambda}(k,q,r) &=& 
{g^2 \over 2}
\int_{\bf p} { \partial f({\bf p}) \over \partial p^\beta } \hat{p}^\mu \;  
\Biggl[ r^\beta {\cal T}^{\nu}(r) {\cal T}^{\lambda}(q)
- k^\beta {\cal T}^{\nu}(k) {\cal T}^{\lambda}(q) 
\Biggr] \; ,
\eqa
which is explicitly
\bqa
\Gamma^{\mu\nu\lambda}(k,q,r) &=& 
{g^2 \over 2}
\int_{\bf p} 
{ \partial f({\bf p}) \over \partial p^\beta } 
\;
\hat{p}^\mu \hat{p}^\nu \hat{p}^\lambda
\left( 
{r^\beta \over \hat{p}\!\cdot\!q\;\hat{p}\!\cdot\!r}
-{k^\beta \over \hat{p}\!\cdot\!k\;\hat{p}\!\cdot\!q}
\right) \; .
\label{3gvertex}
\eqa
For isotropic systems the distribution function only
depends on the length of the three-momentum, $|{\bf p}|=p_0$, 
so that derivative of the distribution function becomes
\bqa
{ \partial f({\bf p}) \over \partial p^\beta } 
&=& { \partial f(p_0) \over \partial p_0 } \delta_{\beta i} \hat{\bf p}^i \, , 
\nonumber \\
&=& { \partial f(p_0) \over \partial p_0 } \left(\delta_{\beta 0} - \hat{p}_\beta\right)  \, ,
\eqa
so that this reduces to the well-known isotropic HTL vertex
\bqa
\Gamma^{\mu\nu\lambda}_{\rm HTL}(k,q,r) &=& 
2 m_\infty^2 
\left\langle
\hat{p}^\mu \hat{p}^\nu \hat{p}^\lambda \left(  
{r^0 \over \hat{p}\!\cdot\!q\;\hat{p}\!\cdot\!r}
-{k^0 \over \hat{p}\!\cdot\!k\;\hat{p}\!\cdot\!q} \right)
\right\rangle_{\hat{\bf p}}
\; .
\eqa

%%%%%%%%%%%%%%%%%%%%%%%%%%%%%%%%%%%%%%%%%%%%%%%%%%%%%%%%%%%%%%%%%%%%%%%%5

\subsection{Four-Gluon Vertex}

%%%%%%%%%%%%%%%%%%%%%%%%%%%%%%%%%%%%%%%%%%%%%%%%%%%%%%%%%%%%%%%%%%%%%%%%5

Similar methods can be used to determine the anisotropic 
four-gluon vertex. The resulting four-gluon vertex
for gluons with outgoing momenta $k$, $q$, $r$, and $s$,
Lorentz indices $\mu$, $\nu$, $\lambda$, and $\sigma$,
and color indices $a$, $b$, $c$, and $d$ is
\bqa
\Gamma^{\mu\nu\lambda\sigma}_{abcd}(k,q,r,s) &=&
2 i g^2 \, (2\pi)^4 \delta^{(4)}(k+q+r+s) \, \mbox{tr}\left[t^a\left(t^bt^ct^d+t^dt^ct^b
\right)\right]{\Gamma}^{\mu\nu\lambda\sigma}(k,q,r,s)
\nonumber
\\
&& \hspace{1cm} + \; 2 \; \mbox{cyclic permutations}\;,
\eqa
where the cyclic permutations are of
$(q,\nu,b)$, $(r,\lambda,c)$, and $(s,\sigma,d)$.
The tensor ${\Gamma}^{\mu\nu\lambda\sigma}(k,q,r,s)$
is defined only for $k+q+r+s=0$:
\bqa
\Gamma^{\mu\nu\lambda\sigma}(k,q,r,s) &=& 
g^2 
\int_{\bf p} 
{ \partial f({\bf p}) \over \partial p^\beta } \; 
\hat{p}^\mu \hat{p}^\nu \hat{p}^\lambda \hat{p}^\sigma  
\left( {k^\beta \over \hat{p}\!\cdot\!k \; \hat{p}\!\cdot\!q \; \hat{p}\!\cdot\!(q+r)}
\right.
\nonumber
\\
&&
\hspace{2cm}
\left.
+{(k+q)^\beta \over \hat{p}\!\cdot\!q\;\hat{p}\!\cdot\!r\;\hat{p}\!\cdot\!(r+s)}
+{(k+q+r)^\beta \over \hat{p}\!\cdot\!r\;\hat{p}\!\cdot\!s\;\hat{p}\!\cdot\!(k+s)}\right) \; .
\label{4gvertex}
\eqa
This tensor is totally symmetric in its four indices and traceless in any
pair of indices, e.g. $g_{\mu\nu}\Gamma^{\mu\nu\lambda\sigma}=0$.
It is even under cyclic or anti-cyclic
permutations of the momenta $k$, $q$, $r$, and $s$.
It satisfies the ``Ward identity''
\bqa
q_{\mu}\Gamma^{\mu\nu\lambda\sigma}(k,q,r,s)=
\Gamma^{\nu\lambda\sigma}(k+q,r,s)
-\Gamma^{\nu\lambda\sigma}(k,r+q,s) \, .
\label{ward-t4}
\eqa
It also reduces to the standard HTL result in the isotropic limit.

\section{Conclusions and Discussions}

In this paper we have shown that the Braaten-Pisarski form 
of the HTL effective action can be straightforwardly extended to systems 
in which the parton distribution functions depend on the 
direction of the three-momentum but are homogeneous in space.  We
have also verified that the same result is obtained using collisionless
gauge-covariant transport theory.  The resulting ``hard-loop'' (HL) effective 
action given by Eq.~(\ref{Saniso}) is manifestly gauge invariant and allows us to easily 
construct all of the $n$-point functions for soft quarks and gluons.  We
have derived explicit expressions for the HL quark-gluon vertex (\ref{vertex}),
the triple-gluon vertex (\ref{3gvertex}), and the four-gluon vertex (\ref{4gvertex}).
By construction these vertices obey the appropriate Ward-Takahashi identities
and reduce to the standard HTL results in the isotropic limit.

We have also discussed the extension of the Taylor-Wong form of the
HTL effective action to anisotropic systems.  In this case the extension
does not seem to be as straightforward because of the presence of terms
which are nontrivial also in the static limit.  This can also be seen
from the explicit expressions for the vertices resulting from the expansion
of the HL effective action.  In the isotropic limit the HTL vertices are
all proportional to the 0-components of the four-momentum flowing through
the vertex so that in the static limit these vertices vanish.  This means
that the static effective potential for QCD contains only bare vertices 
plus electric screening of longitudinal modes coming from the static limit
of $\Pi^{00}$.   In the anisotropic case,
however, even the gluon two-point function has a highly non-trivial static
limit involving three mass scales some of which are imaginary \cite{Romatschke:2003ms}.  
The static limit of the higher gluon $n$-point functions, (\ref{3gvertex}) and 
(\ref{4gvertex}), also appears to be non-trivial since the resulting $n$-point 
functions are no longer simply proportional to the 0-components of the 
four-momentum flowing through them.

The results contained in this paper are relevant to determining the
time scales associated with the possible saturation of soft gluonic 
instabilities.  At the level of the two-point function the static effective 
potential contains terms with a negative curvature due to the presence
of electric and magnetic instabilities.  Depending on the sign of the 
contributions from the higher $n$-point functions these terms could either increase the 
instability or provide for an additional non-abelian saturation of the 
instabilities at some non-vanishing vector potential.  It is interesting 
to note that in relativistically hot QED plasmas the Weibel instability \cite{Weibel:1957}
saturates to a quasi-steady state magnetic Bernstein-Greene-Kruskal wave 
\cite{Bernstein:1957,Berger:1972} which causes a strong residual anisotropy to be 
maintained over rather long time scales compared to the collisional time scale 
\cite{Davidson:1972,Yang:1993}.  It will be interesting to see if an
analogous state exists for anisotropic QCD plasmas. Answering this
question will require a detailed study of the static and quasi-static 
limits of the effective action and associated vertex functions derived in 
this paper.

\section*{Acknowledgments}

The authors would like to thank Peter Arnold for discussions.
M.S. was supported by the Austrian Science Fund Project No. M689.

\section*{Note Added}

In Sect.~II we failed to spell out all the conventions used, and
in fact there is a slight inconsistency with respect to the following sections.

Our metric convention is $(+---)$ throughout. In concordance
with Ref.~\cite{HTLreviews,Blaizot:1993be}, the overall sign
of the hard-loop effective
action $S$ and Lagrangian $\cal L$ in Sect.~II is the one
appropriate for a Euclidean formulation, i.e.~such that
they have to be added to $+{1\over4}\int_x F_{\mu\nu}^a F^{\mu\nu}_a$.
The motivation for this sign choice is that the action is unambiguously
defined in the Wick-rotated Euclidean version, whereas in Minkowski space one
should restrict to the subspace $\cal R$, where all gauge-covariant
line derivatives have vanishing kernel as mentioned at the end of Sect.~III.
If one prefers to write out the action in Minkowski space, i.e.~such
that it is added to $-{1\over4}\int_x F_{\mu\nu}^a F^{\mu\nu}_a$,
one should simply reverse the signs in front of all the $\cal L$'s and $S$'s
in Sect.~II.

Sect.~III in fact uses Minkowski space conventions as the current $j_\mu$
can be unambiguously defined in Minkowski space. The relation
$j^\mu=\delta S/\delta A_\mu$ quoted in Sect.~III assumes that the
sign of $S_{\rm aniso}$ is reversed when switching to
the usual Minkowski space conventions.

%\bibliography{bsample}
%\bibliographystyle{utphys}

%************************************************************************|
% Bibliography
%************************************************************************|

\end{document}